\documentclass{article}
\usepackage{spconf,amsmath,graphicx}

\usepackage{color}
\usepackage{enumitem}
\usepackage{hyperref}
\usepackage{multirow}
\usepackage{etoolbox,siunitx}
\robustify\bfseries
\usepackage{booktabs}
\usepackage{balance}
\usepackage{amssymb}
\usepackage{bm}
\usepackage{arydshln}
\usepackage{caption}
\usepackage{cite}

\let\OLDthebibliography\thebibliography
\renewcommand\thebibliography[1]{
  \OLDthebibliography{#1}
  \setlength{\parskip}{0pt}
  \setlength{\itemsep}{0pt plus 0.3ex}
}
\title{The Cocktail Fork Problem:\\Three-Stem Audio Separation for Real-World Soundtracks}
\name{Darius Petermann$^{1,2}$, Gordon Wichern$^1$, Zhong-Qiu Wang$^1$, Jonathan Le Roux$^1$\thanks{This work was performed while D.~Petermann was an intern at MERL.}}
\address{$^1$Mitsubishi Electric Research Laboratories (MERL), Cambridge, MA, USA\\
$^{2}$Indiana University, Department of Intelligent Systems Engineering, Bloomington, IN, USA
}

\begin{document}
\ninept
\maketitle
\setlength{\abovedisplayskip}{2.5pt}
\setlength{\belowdisplayskip}{2.5pt}

\begin{abstract}

The cocktail party problem aims at isolating any source of interest within a complex acoustic scene,
and has long inspired audio source separation research. Recent efforts have mainly focused on separating speech from noise, speech from speech, musical instruments from each other, or sound events from each other. However, separating an audio mixture (e.g., movie soundtrack) into the three broad categories of speech, music, and sound effects (understood to include ambient noise and natural sound events) has been left largely unexplored, despite a wide range of potential applications. This paper formalizes this task as the \emph{cocktail fork problem}, and presents the Divide and Remaster (DnR) dataset to foster research on this topic. DnR is built from three well-established audio datasets (LibriSpeech, FMA, FSD50k), taking care to reproduce conditions similar to professionally produced content in terms of source overlap and relative loudness, and made available at CD quality. We benchmark standard source separation algorithms on DnR, and further introduce a new multi-resolution model to better address the variety of acoustic characteristics of the three source types. Our best model produces SI-SDR improvements over the mixture of 11.0 dB for music, 11.2 dB for speech, and 10.8 dB for sound effects.

\end{abstract}
\vspace{-.1cm}
\begin{keywords}
audio source separation, speech, music, sound effects, soundtrack
\end{keywords}
\vspace{-.2cm}
\section{Introduction}
\label{sec:intro}
\vspace{-.1cm}

\begin{figure}[t]
    \centering
        \includegraphics[width=.97\linewidth]{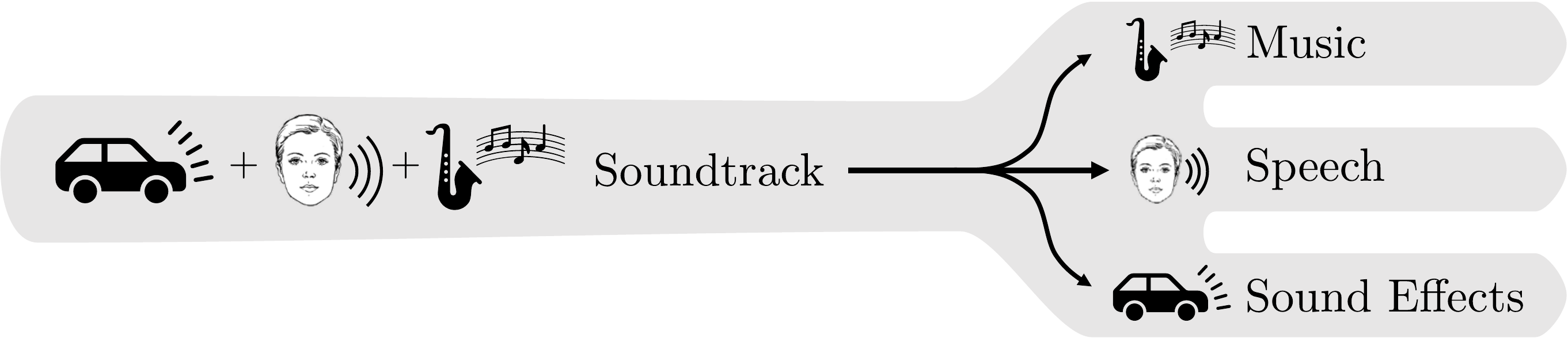}\vspace{-.1cm}
    \caption{Illustration of the cocktail fork problem: given a soundtrack consisting of an audio mixture of speech, music, and sound effects, the goal is to separate it into the three corresponding stems.}\vspace{-.6cm}
    \label{fig:cocktail_fork}
\end{figure}

Humans are able to focus on a source of interest within a complex acoustic scene, a task referred to as the cocktail party problem~\cite{cherry1953some,mcdermott2009cocktail}. %
Research in audio source separation has been dedicated to enabling machines to solve this task, with many studies taking a stab at various slices of the problem, such as the separation of speech from non-speech 
in speech enhancement~\cite{WDL2018, reddy2020interspeech}, speech from other speech in speech separation~\cite{Hershey2016,Drude2019, wichern2019wham},  or separation of individual musical instruments~\cite{rafii2017musdb, stoter19, manilow2019slakh} or non-speech sound events (or sound effects)~\cite{kavalerov2019universal,Tzinis_ICASSP2020, pishdadian2020finding,ochiai2020listen}. %
However, separation of sound mixtures involving speech, music, and sound effects/events has been left largely unexplored, despite its relevance to most produced audio content, such as podcasts, radio broadcasts, and video soundtracks. %
We here intend to bite into this smaller chunk of the cocktail party problem by proposing to separate such soundtracks into these three broad categories. We refer to this task as the \emph{cocktail fork problem}, as illustrated in Fig.~\ref{fig:cocktail_fork}.

While there has been much work on labeling recordings based on these three categories~\cite{theodorou2014overview, melendez2019open, venkatesh2021artificially}, the ability to separate audio signals into these streams has the potential to support a wide range of novel applications.  For example, an end-user could take over the final mixing process by applying independent gains to the separated speech, music, and sound effects signals to support their specific listening environment and preferences.  Furthermore, this three-stream separation could be a front-end for total transcription~\cite{moritz2020all} or audio-visual video description~\cite{hori2017attention} where we want to not only transcribe speech but also semantically describe in great detail the non-speech sounds present in an auditory scene. A recent concurrent work~\cite{zhang2021multitask} also explores the task of speech, music, and sound effects (therein referred to as noise) separation, but only considers the unrealistic case of fully-overlapped mixtures of the three streams, and a low sampling rate of 16 kHz. This sampling rate is not conducive to applications where humans may listen to the separated signals, and it is often difficult or impractical to transition systems trained only on fully-overlapped mixtures to real-world scenarios~\cite{chen2020continuous}.

To provide a realistic high-quality dataset for the cocktail fork problem, we introduce the Divide and Remaster (DnR) dataset, which is built upon LibriSpeech \cite{librispeech_dataset} for speech, Free Music Archive (FMA) \cite{fma_dataset} for music, and Freesound Dataset 50k (FSD50K) \cite{fsd50k_dataset} for sound effects. %
DnR pays particular attention to the mixing process, specifically the relative level of each of the sources and the amount of inter-class overlap, both of which we hope will ease the transition of models trained with DnR to real-world applications. 
Furthermore, DnR includes comprehensive speech, music genre, and sound event annotations, making it potentially useful for research in speech transcription, music classification, sound event detection, and audio segmentation in addition to source separation.

In this paper, we provide a detailed description of the DnR dataset, and benchmark various source separation models. We find the CrossNet unmix (XUMX) architecture~\cite{sawata2021all}, originally proposed for music source separation, also works well for DnR. We further propose a multi-resolution extension of XUMX, to better handle the wide variety of audio characteristics in the sound sources we are trying to separate. We also address several important practical questions often ignored in the source separation literature, such as the impact of sampling rate on model performance, predicted energy in regions where a source should be silent~\cite{schulze2019weakly}, and performance in various overlapping conditions.  
While we only show here objective evaluations based on synthetic data due to the lack of realistic data with stems, we confirmed via informal listening tests that the trained models perform well on real-world soundtracks from YouTube. Our dataset and real-world examples %
are available online.\footnote{\url{cocktail-fork.github.io}}

\section{The Cocktail Fork Problem}
\label{sec:cocktail}
\vspace{-.1cm}

We consider an audio soundtrack $y$ such that 
\begin{equation}
    y=\sum_{j=1}^3 x_j,
\end{equation}
where $x_1$ is the submix containing all music signals, $x_2$ that of all speech signals, and $x_3$ that of all sound effects. We use the term sound effects (SFX) to broadly cover all sources not categorized as speech or music, and choose it over alternatives such as sound events or noise, as the term is especially relevant to our target application where $y$ is a soundtrack. We here define the cocktail fork problem as that of recovering, from the audio soundtrack $y$, its music, speech, and sound effect submixes, as opposed to extracting individual musical instruments, speakers, or sound effects.

Our goal is to train a machine learning model to obtain estimates $\hat{x}_1$, $\hat{x}_2$, and $\hat{x}_3$ of these submixes. We explore two general classes of models for estimating $\hat{x}_j$. The first one, exemplified by Conv-TasNet~\cite{luo2019convTasNet}, takes as input the time-domain mixture $y$, and outputs time-domain estimates $\hat{x}_j$.  The second one operates on the time-frequency (TF) domain mixture, i.e., $Y=\text{STFT}(y)$, and estimates a real-valued mask $\hat{M}_j$ for each source, obtaining time-domain estimates via inverse STFT as $\hat{x}_j=\mathrm{iSTFT}(\hat{M}_j \odot Y)$.  %

\vspace{-.2cm}
\section{Multi-resolution CrossNet (MRX)}
\vspace{-.1cm}
In our benchmark of various network architectures in Section~\ref{sec:results}, we find consistently strong performance from  CrossNet unmix (XUMX)~\cite{sawata2021all}, which uses multiple parameter-less averaging operations
when simultaneously extracting multiple stems (musical instruments in ~\cite{sawata2021all}). XUMX is an STFT masking-based architecture, and choosing appropriate transform parameters is a key design choice.
Longer STFT windows provide better frequency resolution
at the cost of poorer time resolution,
and vice versa for shorter windows.
Mixtures of signals with diverse acoustic characteristics
could thus benefit from multiple STFT resolutions in their TF encoding. Previous research has proven the efficacy of multi-resolution systems for audio-related tasks, such as in the context of speech enhancement~\cite{koizumi2019trainable}, music separation~\cite{grais2018multi}, speech recognition~\cite{toledano2018multi}, and sound event detection ~\cite{benito2021multi}. 
We thus introduce a multi-resolution extension of XUMX which addresses the typical limitations brought by a single-resolution architecture. In \cite{sawata2021all}, the authors show that using multiple parallel branches to process the input
can help in the separation task. 
We here apply this reasoning further towards multiple STFT resolutions. 

\begin{figure}[t]
\centering
    \includegraphics[width=.9\linewidth]{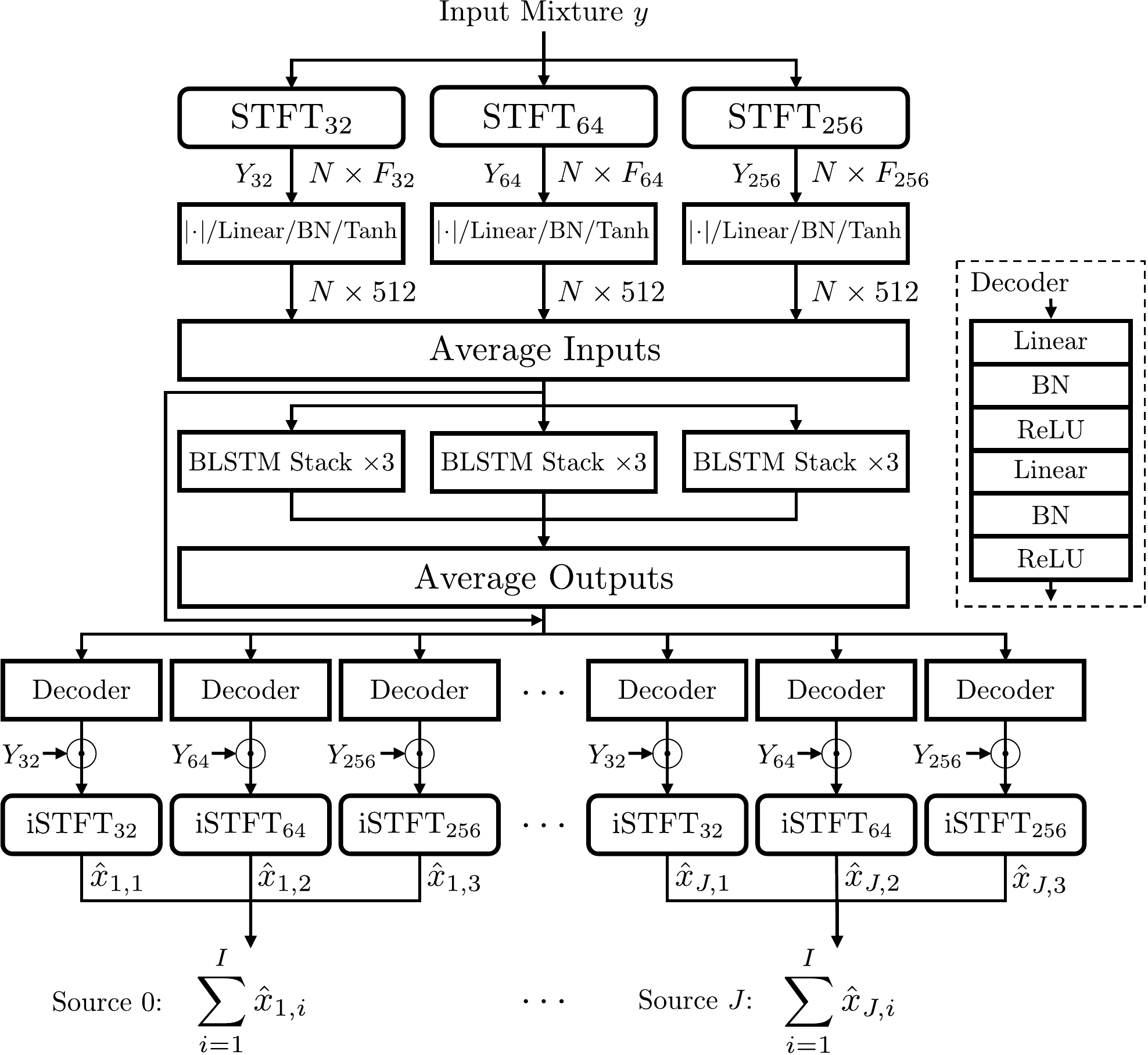}\vspace{-.1cm}
\caption{Multi-resolution CrossNet (MRX) architecture.}\vspace{-.5cm}
\label{fig:xumx_mixed}
\end{figure}

Our proposed architecture %
takes a time-domain input mixture and encodes it into $I$ complex spectrograms $Y_{L_i}$ with different STFT resolutions, where $L_i$ denotes the $i$-th window length in milliseconds. Figure~\ref{fig:xumx_mixed} shows an example with $I=3$ and $\{L_i\}_i=\{32,64,256\}$. %
We use the same hop size (e.g., 8 ms in the example of Fig.~\ref{fig:xumx_mixed}) for all resolutions, so they remain synchronized in time, and $N$ denotes the number of STFT frames for all resolutions. In practice, we set the window size in samples to the nearest power of $2$,
and the number of unique frequency bins is denoted as $F_{L_i}$. 
Each resolution is then passed to a fully connected block to convert the magnitude spectrograms of dimension ${N \times F_{L_i}}$ into a consistent
dimension of $512$ across the resolution branches. This allows us to average them together prior to the bidirectional long short-term memory (BLSTM) stacks, whose outputs are averaged once again. 
While the averaging operators in XUMX were originally intended to efficiently bridge independent architectures for multiple sources, in our case, the input averaging allows the network to efficiently combine inputs with multiple resolutions.

The average inputs and outputs of the BLSTM stacks are concatenated and
decoded back into magnitude soft masks $\hat{M}_{j,i}$, one for each of the three sources $j$ and each of the $I$ original input resolutions $i$. The decoder consists of two stacks of fully-connected layers, each followed by batch normalization (BN) and rectified linear units (ReLU). For a given source $j$, each magnitude mask $\hat{M}_{j,i}$ is multiplied element-wise with the original complex mixture spectrogram  $Y_{L_i}$ for the corresponding resolution, a corresponding time-domain signal $\hat{x}_{j,i}$ is obtained via inverse STFT, and the estimated time-domain signal $\hat{x}_{j}$ is obtained by summing the time-domain signals:
\begin{equation}
\hat{x}_{j} = \sum_{i=1}^{I} \hat{x}_{j,i}= \sum_{i=1}^{I}\text{iSTFT}( \hat{M}_{j,i} \odot Y_{L_i} ).
\end{equation}

For the cocktail fork problem, the network has to estimate a total of $3I$ masks (9 in the example of Fig.~\ref{fig:xumx_mixed}). %
Since ReLU is used as the final mask decoder nonlinearity, the network can freely learn weights for each resolution that best reconstruct the time-domain signal.

\vspace{-.2cm}
\section{DnR Dataset}
\label{sec:dnr}
\vspace{-.1cm}

\subsection{Dataset Building Blocks}
In selecting existing speech, music, and sound effects audio datasets for the cocktail fork problem, we had three primary objectives: (1) the data should be freely available under a Creative Commons license; (2) the sampling rate of the audio should be high enough to cover the full range of human hearing (e.g., 44.1 kHz) to support listening applications (one can always downsample as needed); %
and (3) the audio should contain metadata labels such that it can also be used to explore the impact of separation on downstream tasks, such as transcribing speech and/or providing time-stamped labels for sound effects and music.  We selected the following three datasets.

\noindent {\bf FSD50K - Sound effects:} 
The Freesound Dataset 50k (FSD50K) \cite{fsd50k_dataset} contains 44.1 kHz mono audio, and clips are tagged using a vocabulary of 200 class labels from the AudioSet ontology~\cite{gemmeke2017audioset}. For mixing purposes, we manually classify each of the 200 class labels in FSD50K into one of 3 groups: foreground sounds (e.g., dog bark),  background sounds (e.g., traffic noise), and speech/musical instruments (e.g., guitar, speech).  Speech and musical instrument clips are filtered out to avoid confusion with our speech and music datasets, and we use different mixing rules for foreground and background events as described in Section~\ref{sec:creation_proc}.  We also remove any leading or trailing silence from each sound event prior to mixing.
 
\noindent {\bf Free Music Archive - Music:} 
The Free Music Archive (FMA)~\cite{fma_dataset} is a music dataset including over 100,000 stereo songs across 161 musical genres at 44.1 kHz sampling rate. FMA was originally proposed to address music information retrieval (MIR) tasks and thus includes a wide variety of %
musical metadata. In the context of DnR, we only use track genre as the music metadata.
We use the medium subset of FMA which contains 30 second clips from 25,000 songs in 16 unbalanced genres, and is of a comparable size to FSD50K.
 
\noindent {\bf LibriSpeech - Speech:} 
DnR's speech class is drawn from the LibriSpeech dataset~\cite{librispeech_dataset}, an automatic speech recognition corpus based on public-domain audio books. %
We use the 100 h \textsc{train-clean-100} subset for training, chosen over \textsc{train-clean-360} because it is closer in size to FSD50K and FMA-medium. For validation and test, we use the clean subsets \textsc{dev-clean} and \textsc{test-clean} to avoid noisy speech being confused with music or sound effects. %
We incorporate the provided speech transcription for each utterance as part of the DnR metadata.
LibriSpeech provides its data as clips containing a single speech utterance at 16 kHz. Fortunately, the original 44.1 kHz mp3 audio files containing the unsegmented audiobook recordings harvested from the LibriVox project are also available along with the metadata mapping each LibriSpeech utterance to the original LibriVox filename and corresponding time-stamp, which we use to create a high sampling rate version of LibriSpeech. 

\vspace{-.2cm}
\subsection{Mixing procedure}
\label{sec:creation_proc}
\vspace{-.1cm}

In order to create realistic mixtures of synthetic soundtracks, we focused our effort in two main areas, class overlap and relative level between the different sources in the mixture. Multi-channel spatialization is another important aspect of the mixing process, however, we were unable to find widely agreed upon rules for this process, and therefore focus exclusively on the single-channel case.  We also note that trained single-channel models can be applied independently to each channel of a multi-channel recording, and the outputs combined with a multi-channel Wiener filter for post-processing~\cite{nugraha2016multichannel}. For the purposes of the mixing procedure described in this section, there are four classes: speech, music, foreground effects, and background effects, but the foreground and background sounds are combined into a single submix in the final version of the DnR dataset.

In order to ensure that a mixture could contain multiple full speech utterances and feature a sufficient number of onsets and offsets between the different classes, we decided to make each mixture 60 seconds long.
We do not allow within-class overlap between clips, i.e., two music files will not overlap, but foreground and background sound effects can overlap.  The number of files for each class is sampled %
from a zero-truncated Poisson distribution with expected value $\lambda$. The values of $\lambda$ are chosen based on the average file length of each class, e.g., music and background effects tend to be longer (see Table~\ref{table:creation_params}). For speech files, we always include the entire utterance so that the corresponding transcription remains relevant, while for other classes, we randomly sample the amount of silence between clips of the same class, the clip length, and the internal start time of each clip. Using this mixing procedure, the ``all sources active'' frames account for $\approx 55\%$ of the DnR test set, the ``two sources'' frames for $\approx 32\%$, and ``one source'' frames for $\approx 10\%$, leaving silent frames at $\approx 3\%$ (See Table~\ref{table:overlap_results} for more details).

Regarding the relative amplitude levels across the three classes, after analyzing studies such as~\cite{chaudhuri2018ava} and informal mixing rules from industries such as motion pictures, video games, and podcasting, we found that levels remain fairly consistent across classes, where speech is generally found at the forefront of the mix, followed by foreground sound effects, then music, and finally background ambiances. 
Table~\ref{table:creation_params} depicts the levels used in the DnR dataset in loudness units full-scale (LUFS) \cite{grimm2010lufs}.  To add variability while keeping a realistic consistency over an entire mixture, we first sample an average LUFS value for each class in each mixture, uniformly from a range of $\pm 2.0$ %
around the corresponding Target LUFS.  Then each sound file added to the mix has its individual gain further adjusted by uniformly sampling from a range of $\pm 1.0$.

We base our training, validation, and test splits off of those provided by each of the dataset building blocks. The number of test set mixtures is determined such that we exhaust all utterances from the LibriSpeech \textsc{test-clean} set twice.  We then choose the number of training and validation set mixtures to correspond to a .7/.1/.2 split between training/validation/test, which is roughly in line with the split percentages for FMA (.8/.1/.1) and FSD50k (.7/.1/.2). In the end, DnR consists of 3,406 mixtures ($\approx 57$ h) for the training set, 487 mixtures ($\approx 8$ h) for the validation set, and 973 mixtures ($\approx16$ h) for the test set,
along with their isolated ground-truth stems. %

\begin{table}[t]
\scriptsize
\centering
  \sisetup{table-format=2.1,round-mode=places,round-precision=1,table-number-alignment = center,detect-weight=true,detect-inline-weight=math}
  \caption{Parameters used in the DnR creation procedure.}\label{table:creation_params}\vspace{-.3cm}
\begin{tabular}[t]{lcccc}
\toprule
&{Music}&{Speech}&{SFX-FG}&{SFX-BG}\\
\midrule
$\lambda$ & $\phantom{-2}7\phantom{.0}$ & $\phantom{-1}8\phantom{.0}$ & $\phantom{-}12\phantom{.0}$ & $\phantom{-2}6\phantom{.0}$ \\
Target LUFS & $-24.0$ & $-17.0$ & $-21.0$ &$-29.0$\\
\bottomrule
\end{tabular}\vspace{-.4cm}
\end{table}

 \vspace{-.2cm}
\section{Experimental validation}
\label{sec:experiment}
\vspace{-.1cm}

\subsection{Setup}
\vspace{-.1cm}

We benchmark the performance of several source separation models in terms of scale-invariant signal-to-distortion ratio (SI-SDR)~\cite{leroux2019sdr} for the cocktail fork problem on the DnR dataset, both in the original 44.1 kHz version and in a downsampled 16 kHz version. Unless otherwise noted, we compute the SI-SDR on each 60 second mixture, and average over all tracks in the test set.

\noindent{\bf XUMX and MRX models}:
We consider single-resolution XUMX baselines with various STFT resolutions. We opt to cover a wide range of window lengths $L$ (between 32 and 256 ms) to assess the impact of resolution on performance. %
For our proposed MRX model, we use three STFT resolutions of 32, 64, and 256 ms, which we found to work best on the validation set.
We use XUMX$_\text{L}$ to denote a model with an $L$ ms window. 
We set the hop size to a quarter of the window size.
For the MRX model, %
we determine hop size based on the shortest window. %
To parse the contributions of the multi-resolution and multi-decoder features of MRX, we also evaluate an architecture adding MRX's multi-decoder to the best single-resolution model (XUMX$_\text{64}$), referred to as
XUMX$_\text{64,multi-dec}$. This results in an architecture of the same size (i.e., same number of parameters) as our proposed MRX model. %
In all architectures, each BLSTM layer has 256 hidden units and input/output dimension of 512, and the hidden layer in the decoder has dimension 512.

\noindent{\bf Other benchmarks}: %
We also evaluate our own implementations of Conv-TasNet \cite{luo2019convTasNet} and a temporal convolution network (TCN) with mask inference (MaskTCN). MaskTCN uses an identical TCN to the one used internally by 
Conv-Tasnet, but the learned encoder/decoder are replaced with STFT/iSTFT operations. For MaskTCN, we use an STFT window/hop of 64/16 ms, and for the learned encoder/decoder of Conv-TasNet, we use 500 filters with a window size of 32 samples and a stride of 16 at 16 kHz, and a window size of 80 samples and a stride of 40 at 44.1 kHz. All TCN parameters in both Conv-TasNet and MaskTCN follow the best configuration of~\cite{luo2019convTasNet}.
Additionally, we evaluate Open-Unmix (UMX)~\cite{stoter19}, the predecessor to XUMX, by training a separate model for each source, but without the
parallel branches and averaging operations introduced by XUMX. We also explore a new multi-resolution UMX (MRU), which uses the same settings as the MRX model in Fig.~\ref{fig:xumx_mixed}, but features a single BLSTM stack and a single decoder and is trained separately for each source.

\noindent{\bf Training setup}:
The Conv-TasNet, XUMX$_L$, 
XUMX$_\text{64,multi-dec}$, UMX, MRU, and MRX models all use SI-SDR \cite{luo2019convTasNet,leroux2019sdr} as loss function, while MaskTCN uses the waveform domain $L_1$ loss. %
All models are trained on 9 s chunks, except MaskTCN, trained on 6 s chunks, and Conv-TasNet, trained on 4 s chunks at 16 kHz and 2 s chunks at 44.1 kHz; we found these values to lead to best performance under our GPU memory constraints. All models %
are trained for 300 epochs using ADAM. The learning rate is initialized to $10^{-3}$, and halved if the validation loss is not improved over 3 epochs. %

\vspace{-.2cm}
\subsection{Results and Discussion}
\label{sec:results}
\vspace{-.1cm}

\noindent{\bf Model comparisons}:
Table~\ref{table:model_results} presents the SI-SDR of various models trained and tested on DnR, in addition to the no processing condition (lower bound, using the mixture as estimate) and oracle phase sensitive mask~\cite{erdogan2015psf} (upper bound). 
For each model, SI-SDR improvements are fairly consistent across source types, despite the differences in their relative levels in the mix, which can be seen in the ``No Processing'' SI-SDR.
For both sampling rates, we observe that our proposed MRX model outperforms all single resolution baselines on all source types. %
This implies that the network learns to effectively combine information from different STFT resolutions to more accurately reconstruct the target sources.
The performance of XUMX$_\text{64,multi-dec}$ further confirms this hypothesis by performing nearly identically to XUMX$_\text{64}$, showing that the use of multiple decoders alone does not improve performance.
We also observe that the single-source models (UMX, MRU) tend to perform comparably to the cross-source models (XUMX, MRX) for speech, but perform worse for music and SFX. We speculate that because music and SFX are quieter in the mix, %
it is harder for the network to isolate them effectively without the support of the other sources, while louder sources (here, speech) do not benefit from joint estimation.

\begin{table}[t]
\scriptsize
\centering
  \sisetup{table-format=2.1,round-mode=places,round-precision=1,table-number-alignment = center,detect-weight=true,detect-inline-weight=math}
\caption{
SI-SDR [dB] results of proposed models and baselines on DnR.
}\vspace{-.3cm}
\setlength{\tabcolsep}{3.5pt}
{%
\begin{tabular}[t]{lS[table-format=2.1]S[table-format=2.1]S[table-format=2.1]S[table-format=2.1]S[table-format=2.1]S[table-format=2.1]}
\toprule
&\multicolumn{3}{c}{16 kHz} & \multicolumn{3}{c}{44.1 kHz}\\
\cmidrule(lr){2-4} \cmidrule(lr){5-7} 
\textbf{Model}              & {Music} & {Speech} & {SFX}  & {Music} & {Speech} & {SFX}   \\
\midrule
No processing                           & -6.75 & 1.09   & -5.21        & -6.84 & 0.99   & -4.97  \\ 
Oracle PSF~\cite{erdogan2015psf}        & 11.88 & 17.96   & 13.57       & 11.57 & 17.80   & 13.66  \\ 
\midrule
Conv-TasNet~\cite{luo2019convTasNet}    & 1.14 & 9.43 & 2.74            & 0.28 & 8.48 & 1.96  \\
MaskTCN$_\text{64}$~\cite{luo2019convTasNet} & 1.9 & 10.24 & 3.08       & 1.74 & 9.71 & 3.79 \\
UMX$_\text{64}$~\cite{stoter19}         & 3.17  & 11.76  & 4.12         & 3.05 & 11.74 & 4.37 \\ 
XUMX$_\text{32}$~\cite{sawata2021all}   & 3.15 & 11.46 & 4.47         & 2.86 & 11.2 & 4.73   \\
XUMX$_\text{64}$~\cite{sawata2021all}   & 3.6 & 11.77 & 4.72         & 3.48 & 11.72 & 5.09   \\
XUMX$_\text{128}$~\cite{sawata2021all}  & 3.52 & 11.31 & 4.51          & 3.66 & 11.58 & 5.06 \\
XUMX$_\text{256}$~\cite{sawata2021all}  & 2.93 & 10.17 & 3.8         & 2.94 & 10.52 & 4.35   \\
XUMX$_\text{64,multi-dec}$              & 3.65 & 11.71 & 4.76         & 3.46 & 11.75 & 5.03   \\
MRU$_\text{single-stack}$  (proposed)                          &  3.67   & 12.07 & 4.12      &3.54 & 11.8  &4.39 \\
MRU$_\text{multi-stack}$  (proposed)                          &  3.73   & \bfseries 12.2 & 4.15      &3.76 & 11.95  &4.6 \\
MRX (proposed)                          &  \bfseries 4.08 &  \bfseries 12.21 &  \bfseries 5.1   &  \bfseries 4.21 &  \bfseries 12.33 &  \bfseries 5.69   \\
\bottomrule
\end{tabular}\vspace{-.5cm}
\label{table:model_results}
}
\end{table}

\noindent{\bf Sampling rate comparisons}:
Table \ref{table:sr_results} compares the average SI-SDR performance of the MRX model across sampling rates. In the 44.1 kHz column, we observe a reduction in SI-SDR for all source classes when upsampling the 16 kHz model output to 44.1 kHz (``Resampled'').  This is to be expected as all frequencies above 8 kHz are zero in the upsampled 44.1 kHz signal, but we see that these frequencies only contribute a small amount to the difference in SI-SDR scores, as there is comparatively little energy there. %
In the 16 kHz column, we observe a minor performance gain in the ``Resampled'' row where the 44.1 kHz model output is downsampled to 16 kHz, showing that the model can make use of information above 8 kHz to improve separation under 8 kHz.  This result could be beneficial for transcription applications where ASR or sound event detection models are pre-trained at 16 kHz, but a front-end source separation model can obtain better separated signals using 44.1 kHz input. 

\begin{table}[t]
\scriptsize
\centering
  \sisetup{table-format=2.1,round-mode=places,round-precision=1,table-number-alignment = center,detect-weight=true,detect-inline-weight=math}
\caption{%
SI-SDR [dB] of the MRX model across sample rates. In Original row, processing is done at the evaluation sample rate indicated in each column. In Resampled row, processing is done at the other sample rate (44.1 kHz for 16 kHz and vice versa) and the output resampled for evaluation.}
\vspace{-.3cm}
\setlength{\tabcolsep}{3.5pt}
{%
\begin{tabular}[t]{lS[table-format=1.1]S[table-format=2.1]S[table-format=1.1]S[table-format=1.1]S[table-format=2.1]S[table-format=1.1]}
\toprule
&\multicolumn{3}{c}{16 kHz} & \multicolumn{3}{c}{44.1 kHz}\\
\cmidrule(lr){2-4} \cmidrule(lr){5-7} 
\textbf{Model}              & {Music} & {Speech} & {SFX}  & {Music} & {Speech} & {SFX}   \\
\midrule
Original                      & 4.08  & 12.22  & 5.09  & 4.24  & 12.32  & 5.68 \\
Resampled                     & 4.24  & 12.34  & 5.44   & 3.94  & 11.25  & 4.12  \\

\bottomrule
\end{tabular}\vspace{-.3cm}
\label{table:sr_results}
}
\end{table}

\begin{table}[t]
    \scriptsize
    \centering
    \sisetup{table-format=2.1,round-mode=places,round-precision=1,table-number-alignment = center,detect-weight=true,detect-inline-weight=math}
    \caption{Performance of MRX at 44.1 kHz for each source on the seven overlapping use-cases. M, S, X indicate presence within a frame of music, speech, and SFX respectively. Scores indicate SI-SDR [dB] improvement when the source is present with another, final SI-SDR [dB] when the source is the only one present (indicated by $*$), and Predicted Energy at Silence [dB] (PES) when a source is not present (indicated by $\textbf{\dag}$).}\vspace{-.3cm}
    \setlength\tabcolsep{2.0pt}
    \captionsetup[table]{skip=5pt}
    \resizebox{\linewidth}{!}{
    \begin{tabular}{lSSSSSSS}
    \toprule
    \textbf{Source} & \{{M,S,X}\} & \{{$\emptyset$,S,X}\}  & \{{M,$\emptyset$,X}\}  & \{{M,S,$\emptyset$}\}  & \{{M,$\emptyset$,$\emptyset$}\} & \{{$\emptyset$,S,$\emptyset$}\} & \{{$\emptyset$,$\emptyset$,X}\} \\
    \textbf{Frames} & {$31910$} & \num{6351} & \num{8702} & \num{3761} & \num{1330} & \num{1761} & \num{2737} \\
    \midrule
        Music      & 10.22 & -13.64$^\textbf{\dag}$ & 2.9 & 8.97 & 10.74$^*$ & -23.3$^\textbf{\dag}$ & -21.55$^\textbf{\dag}$ \\
        Speech     & 11.78 & 8.99 & -22.024$^\textbf{\dag}$ & 10.43 & -28.654$^\textbf{\dag}$ & 23.39$^*$ & -29.024$^\textbf{\dag}$           \\
        SFX         & 12.03 & 10.79 & 7.22 & -6.78$^\textbf{\dag}$ & -11.12$^\textbf{\dag}$ & -13.38$^\textbf{\dag}$ & 11.6$^*$\\
    \bottomrule
    \end{tabular}}\vspace{-.5cm}
    \label{table:overlap_results}
\end{table}

\noindent{\bf Overlap scenario comparisons}: 
We here compute metrics over 1 s segments to evaluate performance independently in regions where only certain source classes overlap. Metrics %
such as SI-SDR or SDR \cite{vincent2006bss}
are undefined for signals containing silent target and/or estimated sources. This limitation is usually circumvented by disregarding the %
problematic frames in the evaluation process \cite{stoter19} (i.e., frames containing one or more silent target(s) and/or estimated source(s)). For example, in the MUSDB test set~\cite{rafii2017musdb}, it is reported that at least 45 minutes out of a total of 210 minutes of test data are systematically ignored for that reason~\cite{schulze2019weakly}. 
Although these regions with fewer active sources may be seen as less challenging, we believe it is important to also evaluate performance when not all sources are present, and here consider all types of overlap. 

Table \ref{table:overlap_results} shows the overall results for each of the three sources in the seven possible overlapping scenarios.  For regions where a source is not active in the ground truth, we report predicted energy at silence (PES)~\cite{schulze2019weakly} to quantify the energy incorrectly assigned to a silent source. We note that speech has smaller PES values than music or SFX in Table~\ref{table:overlap_results}, even though it is the loudest source on average.
SI-SDR in the single-source cases are very high, especially for speech, showing that few artifacts are introduced. Among the two-source cases, we note that SI-SDRi is substantially lower for music and SFX (\{{M,$\emptyset$,X}\}), indicating that these two sets of varied sources are more difficult to separate from each other than from speech.

\vspace{-.3cm}
\section{Conclusion}
\label{sec:conclusion}
\vspace{-.2cm}

In this paper, we formalized the task of three-stem soundtrack separation as the cocktail fork problem, and introduced DnR, a high-quality dataset built on top of three well-established sound collections: LibriSpeech (speech), FSD50K (SFX), and FMA (music). %
We benchmarked several source separation algorithms on DnR  and 
showed that our proposed multi-resolution model performed best. 
In the future, we plan to combine the separation models developed in this paper with speech recognition and sound classification systems for automatic caption generation of speech and non-speech sounds, and explore remixing strategies that minimize perceptual artifacts~\cite{torcoli2021controlling}.

\bibliographystyle{IEEEtran}
\bibliography{refs}

\begin{thebibliography}{10}
\providecommand{\url}[1]{#1}
\csname url@samestyle\endcsname
\providecommand{\newblock}{\relax}
\providecommand{\bibinfo}[2]{#2}
\providecommand{\BIBentrySTDinterwordspacing}{\spaceskip=0pt\relax}
\providecommand{\BIBentryALTinterwordstretchfactor}{4}
\providecommand{\BIBentryALTinterwordspacing}{\spaceskip=\fontdimen2\font plus
\BIBentryALTinterwordstretchfactor\fontdimen3\font minus
  \fontdimen4\font\relax}
\providecommand{\BIBforeignlanguage}[2]{{%
\expandafter\ifx\csname l@#1\endcsname\relax
\typeout{** WARNING: IEEEtran.bst: No hyphenation pattern has been}%
\typeout{** loaded for the language `#1'. Using the pattern for}%
\typeout{** the default language instead.}%
\else
\language=\csname l@#1\endcsname
\fi
#2}}
\providecommand{\BIBdecl}{\relax}
\BIBdecl

\bibitem{cherry1953some}
E.~C. Cherry, ``Some experiments on the recognition of speech, with one and
  with two ears,'' \emph{J. Acoust. Soc. Am.}, vol.~25, no.~5, pp. 975--979,
  1953.

\bibitem{mcdermott2009cocktail}
J.~H. McDermott, ``The cocktail party problem,'' \emph{Current Biology},
  vol.~19, no.~22, pp. R1024--R1027, 2009.

\bibitem{WDL2018}
D.~Wang and J.~Chen, ``Supervised speech separation based on deep learning: An
  overview,'' \emph{IEEE/ACM Trans. Audio, Speech, Lang. Process.}, vol.~26,
  no.~10, pp. 1702--1726, 2018.

\bibitem{reddy2020interspeech}
C.~K. Reddy, V.~Gopal, R.~Cutler, E.~Beyrami \emph{et~al.}, ``The {I}nterspeech
  2020 {D}eep {N}oise {S}uppression challenge: Datasets, subjective testing
  framework, and challenge results,'' in \emph{Proc. Interspeech}, Oct. 2020.

\bibitem{Hershey2016}
J.~R. Hershey, Z.~Chen, and J.~Le~Roux, ``Deep clustering: Discriminative
  embeddings for segmentation and separation,'' in \emph{Proc. ICASSP}, Mar.
  2016, pp. 31--35.

\bibitem{Drude2019}
L.~Drude, J.~Heitkaemper, C.~Boeddeker, and R.~Haeb-Umbach, ``{SMS-WSJ}:
  Database, performance measures, and baseline recipe for multi-channel source
  separation and recognition,'' in \emph{arXiv preprint arXiv:1910.13934},
  2019.

\bibitem{wichern2019wham}
G.~Wichern, J.~Antognini, M.~Flynn, L.~R. Zhu \emph{et~al.}, ``{WHAM}!:
  Extending speech separation to noisy environments,'' in \emph{Proc.
  Interspeech}, Sep. 2019.

\bibitem{rafii2017musdb}
\BIBentryALTinterwordspacing
Z.~Rafii, A.~Liutkus, F.-R. St{\"o}ter, S.~I. Mimilakis \emph{et~al.}, ``The
  {MUSDB18} corpus for music separation,'' Dec. 2017. [Online]. Available:
  \url{https://doi.org/10.5281/zenodo.1117372}
\BIBentrySTDinterwordspacing

\bibitem{stoter19}
F.-R. St{\"o}ter, S.~Uhlich, A.~Liutkus, and Y.~Mitsufuji, ``Open-{U}nmix - a
  reference implementation for music source separation,'' \emph{Journal of Open
  Source Software}, 2019.

\bibitem{manilow2019slakh}
E.~Manilow, G.~Wichern, P.~Seetharaman, and J.~Le~Roux, ``Cutting music source
  separation some {Slakh}: A dataset to study the impact of training data
  quality and quantity,'' in \emph{Proc. WASPAA}, Oct. 2019.

\bibitem{kavalerov2019universal}
I.~Kavalerov, S.~Wisdom, H.~Erdogan, B.~Patton \emph{et~al.}, ``Universal sound
  separation,'' in \emph{Proc. WASPAA}, Oct. 2019.

\bibitem{Tzinis_ICASSP2020}
E.~Tzinis, S.~Wisdom, J.~R. Hershey, A.~Jansen \emph{et~al.}, ``Improving
  universal sound separation using sound classification,'' in \emph{Proc.
  ICASSP}, May 2020.

\bibitem{pishdadian2020finding}
F.~Pishdadian, G.~Wichern, and J.~Le~Roux, ``Finding strength in weakness:
  Learning to separate sounds with weak supervision,'' \emph{IEEE/ACM Trans.
  Audio, Speech, Lang. Process.}, vol.~28, pp. 2386--2399, 2020.

\bibitem{ochiai2020listen}
T.~Ochiai, M.~Delcroix, Y.~Koizumi, H.~Ito \emph{et~al.}, ``Listen to what you
  want: Neural network-based universal sound selector,'' in \emph{Proc.
  Interspeech}, Oct. 2020.

\bibitem{theodorou2014overview}
T.~Theodorou, I.~Mporas, and N.~Fakotakis, ``An overview of automatic audio
  segmentation,'' \emph{IJITCS}, vol.~6, no.~11, 2014.

\bibitem{melendez2019open}
B.~Mel{\'e}ndez-Catal{\'a}n, E.~Molina, and E.~G{\'o}mez, ``Open broadcast
  media audio from tv: A dataset of tv broadcast audio with relative music
  loudness annotations,'' \emph{Trans. ISMIR}, vol.~2, no.~1, 2019.

\bibitem{venkatesh2021artificially}
S.~Venkatesh, D.~Moffat, A.~Kirke, G.~Shakeri \emph{et~al.}, ``Artificially
  synthesising data for audio classification and segmentation to improve speech
  and music detection in radio broadcast,'' in \emph{Proc. ICASSP}, Jun. 2021,
  pp. 636--640.

\bibitem{moritz2020all}
N.~Moritz, G.~Wichern, T.~Hori, and J.~Le~Roux, ``All-in-one transformer:
  Unifying speech recognition, audio tagging, and event detection.'' in
  \emph{Proc. Interspeech}, Oct. 2020.

\bibitem{hori2017attention}
C.~Hori, T.~Hori, T.-Y. Lee, Z.~Zhang \emph{et~al.}, ``Attention-based
  multimodal fusion for video description,'' in \emph{Proc. CVPR}, Jul. 2017.

\bibitem{zhang2021multitask}
L.~Zhang, C.~Li, F.~Deng, and X.~Wang, ``Multi-task audio source separation,''
  \emph{arXiv preprint arXiv:2107.06467}, 2021.

\bibitem{chen2020continuous}
Z.~Chen, T.~Yoshioka, L.~Lu, T.~Zhou \emph{et~al.}, ``Continuous speech
  separation: Dataset and analysis,'' in \emph{Proc. ICASSP}, May 2020, pp.
  7284--7288.

\bibitem{librispeech_dataset}
V.~Panayotov, G.~Chen, D.~Povey, and S.~Khudanpur, ``Librispeech: An {ASR}
  corpus based on public domain audio books,'' in \emph{Proc. ICASSP}, Apr.
  2015, pp. 5206--5210.

\bibitem{fma_dataset}
M.~Defferrard, K.~Benzi, P.~Vandergheynst, and X.~Bresson, ``{FMA}: A dataset
  for music analysis,'' in \emph{Proc. ISMIR}, Oct. 2017.

\bibitem{fsd50k_dataset}
E.~Fonseca, X.~Favory, J.~Pons, F.~Font \emph{et~al.}, ``{FSD50K}: An open
  dataset of human-labeled sound events,'' \emph{arXiv preprint
  arXiv:2010.00475}, 2020.

\bibitem{sawata2021all}
R.~Sawata, S.~Uhlich, S.~Takahashi, and Y.~Mitsufuji, ``All for one and one for
  all: Improving music separation by bridging networks,'' in \emph{Proc.
  ICASSP}, Jun. 2021, pp. 51--55.

\bibitem{schulze2019weakly}
K.~Schulze-Forster, C.~Doire, G.~Richard, and R.~Badeau, ``Weakly informed
  audio source separation,'' in \emph{Proc. WASPAA}, Oct. 2019.

\bibitem{luo2019convTasNet}
Y.~Luo and N.~Mesgarani, ``Conv-{T}as{N}et: Surpassing ideal time-frequency
  magnitude masking for speech separation,'' \emph{IEEE/ACM Trans. Audio,
  Speech, Lang. Process.}, vol.~27, no.~8, pp. 1256--1266, 2019.

\bibitem{koizumi2019trainable}
Y.~Koizumi, N.~Harada, and Y.~Haneda, ``Trainable adaptive window switching for
  speech enhancement,'' in \emph{Proc. ICASSP}, May 2019.

\bibitem{grais2018multi}
E.~M. Grais, H.~Wierstorf, D.~Ward, and M.~D. Plumbley, ``Multi-resolution
  fully convolutional neural networks for monaural audio source separation,''
  in \emph{Proc. LVA}, Jul. 2018.

\bibitem{toledano2018multi}
D.~T. Toledano, M.~P. Fern{\'a}ndez-Gallego, and A.~Lozano-Diez,
  ``Multi-resolution speech analysis for automatic speech recognition using
  deep neural networks: Experiments on {TIMIT},'' \emph{PLoS ONE}, vol.~13,
  2018.

\bibitem{benito2021multi}
D.~De~Benito-Gorrón, D.~Ramos, and D.~T. Toledano, ``A multi-resolution
  {CRNN}-based approach for semi-supervised sound event detection in {DCASE}
  2020 challenge,'' \emph{IEEE Access}, vol.~9, pp. 89\,029--89\,042, 2021.

\bibitem{gemmeke2017audioset}
J.~F. Gemmeke, D.~P. Ellis, D.~Freedman, A.~Jansen \emph{et~al.}, ``Audio set:
  An ontology and human-labeled dataset for audio events,'' in \emph{Proc.
  ICASSP}, Mar. 2017.

\bibitem{nugraha2016multichannel}
A.~A. Nugraha, A.~Liutkus, and E.~Vincent, ``Multichannel audio source
  separation with deep neural networks,'' \emph{IEEE/ACM Trans. Audio, Speech,
  Lang. Process.}, vol.~24, no.~9, 2016.

\bibitem{chaudhuri2018ava}
S.~Chaudhuri, J.~Roth, D.~P. Ellis, A.~Gallagher \emph{et~al.}, ``Ava-speech: A
  densely labeled dataset of speech activity in movies,'' in \emph{Proc.
  Interspeech}, Sep. 2018.

\bibitem{grimm2010lufs}
E.~Grimm, R.~Van~Everdingen, and M.~J. L.~C. Schöpping, ``Toward a
  recommendation for a european standard of peak and {LKFS} loudness levels,''
  \emph{SMPTE Motion Imaging Journal}, vol. 119, no.~3, pp. 28--34, 2010.

\bibitem{leroux2019sdr}
J.~Le~Roux, S.~Wisdom, H.~Erdogan, and J.~R. Hershey, ``{SDR}--half-baked or
  well done?'' in \emph{Proc. ICASSP}, May 2019.

\bibitem{erdogan2015psf}
H.~Erdogan, J.~Hershey, S.~Watanabe, and J.~Le~Roux, ``Phase-sensitive and
  recognition-boosted speech separation using deep recurrent neural networks,''
  in \emph{Proc. ICASSP}, Apr. 2015.

\bibitem{vincent2006bss}
E.~Vincent, R.~Gribonval, and C.~Fevotte, ``Performance measurement in blind
  audio source separation,'' \emph{IEEE Trans. Audio, Speech, Lang. Process.},
  vol.~14, no.~4, pp. 1462--1469, 2006.

\bibitem{torcoli2021controlling}
M.~Torcoli, J.~Paulus, T.~Kastner, and C.~Uhle, ``Controlling the remixing of
  separated dialogue with a non-intrusive quality estimate,'' in \emph{Proc.
  WASPAA}, Oct. 2021.

\end{thebibliography}

\end{document}